\documentclass[12pt,letterpaper]{article}
\usepackage{amsmath,amssymb,pgf,pgfarrows,pgfnodes,float,appendix, hyperref}
\usepackage{graphicx}
\usepackage{subfigure}
\usepackage[margin=0.9in]{geometry}

\newcommand{\be}{\begin{equation}}
\newcommand{\ee}{\end{equation}}
\newcommand{\bea}{\begin{eqnarray}}
\newcommand{\eea}{\end{eqnarray}}

\title{{\rm\footnotesize \qquad \qquad \qquad \qquad \qquad \ \qquad \qquad \qquad \ \ \ \ \ \                  RUNHETC-2015-8     
SCIPP 15/07}\vskip.5in     The Temperature/Entropy Connection for Horizons, Massless Particle Scattering, and the Origin of Locality\\
Essay Written for the Gravitation Research Foundation 2015 Awards for Essays on Gravitation\\ March 20, 2015}
\author{Tom Banks\\
Department of Physics and SCIPP\\
University of California, Santa Cruz, CA 95064\\
{\it and}\\
Department of Physics and NHETC\\
Rutgers University, Piscataway, NJ 08854\\
E-mail: \href{mailto:banks@scipp.ucsc.edu}{banks@scipp.ucsc.edu}
\\
\\
}
\date{}
\begin{document}
\maketitle

\begin{abstract}
 I explain, in non-technical terms, the basic ideas of Holographic Space-time (HST) models of quantum gravity (QG).  The key feature is that the degrees of freedom (DOF) of QG, localized in a finite causal diamond are restrictions of an algebra of asymptotic currents, describing flows of quantum numbers out to null infinity in Minkowski space, with zero energy density on the sphere at infinity.  Finite energy density states are constrained states of these DOF and the resulting relation between asymptotic energy and the number of constraints, explains the relation between black hole entropy and energy, as well as the critical energy/impact parameter regime in which particle scattering leads to black hole formation.  The results of a general class of models, implementing these principles, are described, and applied to understand the firewall paradox, and to construct a finite model of the early universe, which implements inflation with only the minimal fine tuning needed to obtain a universe containing localized excitations more complex than large black holes.
\end{abstract}

\section{Introduction}
Perhaps no recent question in theoretical physics has generated as much confusion as the conundrum of the degrees of freedom (DOF) responsible for the entropy of horizons, and their relation to the the familiar degrees of freedom of Effective Quantum Field Theory (QUEFT).  The purpose of this essay is to explain, in the simplest way I can, the answer to this question proposed in the formalism of Holographic Space-time (HST), which Willy Fischler and I have tried to construct over the past decade and a half.  I'll include almost none of the formal structure of that theory, and refer you instead to our extensive collection of papers\cite{holo}.

Let me make the essential points from the beginning:  Jacobson\cite{ted} showed that, apart from the cosmological constant, Einstein's field equations could be derived as the hydrodynamics of a quantum system whose Hilbert space, in a finite causal diamond, has an entropy equal to one quarter of the area in Planck units, of the diamond's holographic screen.  The holographic screen is the maximal area two surface found on the boundary of the diamond. 
String theory teaches us that all fields in nature are related to the metric of an eleven dimensional space-time or its supersymmetric partners. One only quantizes the equations of hydrodynamics when studying low energy, low entropy excitations of the ground state of a system.

As a consequence, QU(antum) E(ffective) F(ield) T(heory) cannot give a microscopic description of horizon entropy.  The familiar failure to account even parametrically ($R^{3/2}$ {\it vs.} $R^2$) for the entropy in terms of high entropy QUEFT states is merely a symptom of this. QUEFT is an inadequate approximation to the HST theory in high entropy situations which involve the micro-physics of horizons.  The HST theory of the very early universe is also a high entropy regime.

The true DOF of QG {\it can} be understood in QUEFT/particle theory language, in the following manner.  QUEFT can be derived\cite{wein} from the assumption that the low energy degrees of freedom are exhausted by particles, and that the S-matrix obeys the constraints of locality and Lorentz invariance in Minkowski space-time.  In dimension $\geq 4$, in models including gravity, there is a problem with this derivation\footnote{In dimension $< 4$ the derivation fails entirely.  There is no theory of gravitational scattering in Minkowsi space in low dimensions.  The HST formalism gives a natural explanation for this\cite{tbtbpub}.} because such models contain massless particles.  There are incoming and outgoing states in which a finite amount of energy penetrates the holoscreen at infinity, but the local density of energy penetrating that screen vanishes. We will call these states, and their analogs in finite causal diamonds, {\it horizon states}.  In low enough dimension, neglect of this phenomenon leads to infrared divergences in the conventional S-matrix. Energy transfer via horizon states is unavoidable, and IR divergences are the perturbative symptom of the vanishing of amplitudes in which no such transfer occurs. However, processes involving horizon states have non-zero amplitudes in {\it any dimension}, and a definition of scattering, which does not take them into account, will not have a unitary scattering operator.

To address this problem, one is led to define scattering in terms of an algebra of asymptotic currents defined on the holoscreen at infinity\cite{superBMS}\cite{tbtbpub}\cite{andy}. The currents describe transport of quantum numbers, whether by finite energy density particles or zero energy density {\it horizon DOF}, through the holographic screen.  The scattering operator intertwines representations of the current algebra at past and future holoscreens. It is the analog of Christodoulu-Klainerman boundary conditions in GR\cite{CK}\cite{andy}.

Transport of momentum is built in to the definitions of the current algebra. The conformal boundary of Minkowski space is dual to the null momentum light cone, $P^2 = 0$.  Each point on this cone represents a null momentum coming into or out of the sphere at infinity. Arguments too long to present here indicate that the current algebra must contain spinor currents $\psi_{\alpha} (P, p).$\footnote{The finite set of labels $p$ represent internal quantum numbers.} The spinors satisfy the Cartan-Penrose equation $$ P_{\mu} \gamma^{\mu}_{{\alpha}\beta} \psi_{\beta} (P, p ) = 0, $$ which means they're spinors on the two-sphere at infinity. Particle (more properly {\it jet}) representations of the current algebra have $\psi_{\alpha}) (P \neq 0)$ vanishing outside of a finite number of  cones of finite opening angle on the sphere\cite{stermanweinberg}. $\psi_{\alpha} ( P = 0) $ can be non-vanishing everywhere except in small annuli surrounding the openings of the cones.  

Aspects of this general prescription are valid for gauge invariant bulk field theories not including gravitation, as well as, in HST, for a model of quantum gravity in Minkowski space.  The difference between the two types of systems becomes apparent only when we back off from the conformal boundary and consider a finite causal diamond.  In quantum field theory the Hilbert space remains infinite dimensional and there is a sense in which ``most" of the DOF live in the bulk\footnote{To make this more precise we have to introduce an ultraviolet cutoff, and restrict attention to states below some finite energy.  Then, for diamonds whose size is large in cutoff units, we have the usual dominance of bulk over boundary DOF.}.   The model contains mathematical observables beyond those encoded in the S-matrix.   In HST, we find instead that the variables describing the finite diamond are a subset of those describing the infinite diamond.  This is the content of the Covariant Entropy Principle\cite{fsb} which, when taken together with Jacobson's observation\cite{ted} that the entropy in question is that of the infinite temperature Unruh density matrix, tells us that the logarithm of the dimension of the Hilbert space of the diamond is one quarter of the area of its holoscreen, in Planck units. A natural way to impose this constraint is to postulate that the algebra of boundary currents is truncated beyond the spin $N$ terms of its spherical harmonic expansion.  If the radius of the holoscreen is $R$ then $\pi (RM_P)^2 \sim L N^2 $, where $e^L$ is the dimension of the representation of the current algebra for fixed angular momentum.

If we now contemplate the analog of the jet state constraint, for a large causal diamond of holoscreen area $4\pi R^2$ , then the number of constraints we must enforce to ``set the $P = 0$ currents to zero in an annulus" scales like $R$\footnote{Everything we're saying apart from the remarks on de Sitter space and cosmology, can be generalized to higher dimensions\cite{holonewton}.}. The fuzzy sphere cut-off appropriate to this finite area screen is to take the spinors to be $N \times N+1$ matrices (and their adjoints)\cite{tbjk}, which includes all half odd integer angular momenta up to $N - 1/2$. We quantize these variables so that they form the fermionic generators of a super-algebra, whose representation space is swept out by the action of the fermionic generators. This imposes the covariant entropy principle. The constraints correspond to setting $NE + Q$ matrix elements of the square $\psi \psi^{\dagger}$ bilinears equal to zero on the incoming states.  $Q$ is a number of order $1$ as $N \rightarrow \infty$ , which can change from initial to final state.  However, there is a large class of Hamiltonians for which $E$ is an asymptotically conserved quantum number, which we identify with energy, in Planck units.  The constraints allow us to write the square matrices in block diagonal form, with one block of size $o(N)$ and the others finite. A Hamiltonian built from a single trace of a function of the square matrices will have no interaction between blocks.  Fixed states of the variables in the blocks correspond to free particles.

Consider a nested set of causal diamonds, corresponding to a nested set of intervals along a time-like trajectory.  The full Hilbert space of the system has an entropy given by the area of the maximal causal diamond, reached in an infinitely long time interval. The smaller diamonds are proper tensor factors of this Hilbert space. Causality implies that the Hamiltonian propagating the system in proper time along this trajectory, must decouple the DOF inside some small diamond, from those which commute with them\footnote{In actual models, the fundamental variables are the fermionic generators of a super-algebra, and commutativity is really graded commutativity.} in the full Hilbert space.  This means that the Hamiltonian {\it must} be time dependent.  Note that this is in accord with the rule of classical general relativity, that conservation laws are associated with {\it asymptotic symmetries}, which can't be seen in a finite causal diamond.  $K$ above is an example of such an asymptotic symmetry.

The time dependence of the Hamiltonian can be chosen in such a way that, as $N = RM_P \rightarrow \infty$, the number of constraints cannot change by amounts of order $R$.  In this limit, the subsets of DOF we've associated with localized particles, decouple both from the horizon and from each other.  Their energy, as defined above, is conserved in this limit.  The slowest falloff of interactions compatible with this decoupling gives a large impact parameter scattering amplitude that scales with energy and impact parameter like the Newton potential\cite{holonewton}.  Faster falloff of interactions would lead to more asymptotic conservation laws, corresponding to powers of the energy, and this is incompatible with a non-trivial Poincare invariant S-matrix.  Most of our HST models do not give a Poincare invariant S-matrix, but they contain enough parameters that one can conjecture that Poincare invariance can be obtained by tuning\footnote{In fact, this is required by the rules of HST.  Those rules include the consistency of the S-matrix computed along different time-like trajectories, including those related by Poincare transformations.  The models constructed so far are compatible with this rule only for geodesics related by translations and rotations, but not boosts.}.

Consider a fixed asymptotic state, consisting of a finite number of incoming jets, and a fixed time-like trajectory in Minkowski space.  We choose the nesting of causal diamonds to be time symmetric around some point $P$ on the trajectory, which we label $t = 0$.
The Hamiltonian at proper time $t$ is \begin{equation} H(t) = H( - t) = H_{in} (t) + H_{out} (t) . \end{equation}  $H_{in} (t)$ acts on a tensor factor of entropy $\sim t^{2}$ in the full Hilbert space, whose entropy is $\sim T^2$, with $T$ taken to infinity at the end of the calculation.  The incoming state has total energy $E = \sum E_i $, which implies $\sim ET$ degrees of freedom vanish on that state. 

This connection between energy and the number of constraints is crucial to everything that goes on in gravitational physics.  Suppose some fraction $E^{\prime}$ of the constraints refer to DOF which remain in the Hilbert space of the causal diamond of some particular observer down to some small causal diamond of size $R_S$.  If $E^{\prime} \ll R_S$ in Planck units, we will see an initial state in this diamond which looks like a fuzzy version of the jet state.  There's a finite probability that the outgoing state in this diamond will look like a, generally different jet state.  When $E^{\prime} \sim R_S$ this is no longer possible and all the DOF in the small causal diamond will thermalize.  This unconstrained thermal state will equilibrate, and this is what we think of as a black hole.  It has no local excitations inside the diamond (no constraints) after equilibration.  The equilibrium is not stable however, since there's a probability $e^{- R_S \epsilon}$ to accidentally come to a state with $R_S \epsilon$ constraints.  So the black hole will emit particles thermally, with $T_{Hawking} \sim R_S^{-1}$.  This leads to a lifetime of order $R_S^3$.  

If $R_S$ is not too large in Planck units, or we never get to a situation where the small causal diamond has $E^{\prime} \sim R_S$, all of this will look like a particle vertex localized within a few Planck lengths.  It's only when $E^{\prime} \sim R_S \gg 1$ that we see semi-classical black hole production.  All of these models then, contain particle-like excitations, which scatter in a manner consistent with causality, quasi-locality of interaction\footnote{Locality comes from the time dependence of the Hamiltonian, which says that interaction vertices are strong only in small diamonds.}, rotation and translation invariance.
Translation invariance is imposed by the consistency conditions between trajectories related by a space-time translation.  As shown in \cite{fw3} this leads to a Feynman diagram classification of amplitudes.  The models also have a quite explicit quantum mechanical picture of black hole formation and evaporation.

The same set of ideas determine the temperature and entropy of dS space. In the limit of large de Sitter radius, our proposed model of eternal dS space is simply the model of Minkowski space, with $R$, the area of the maximal holographic screen, left finite, while the proper time $t$ can go to infinity. For $ t > R$ the Hamiltonian is chosen to be constant and equal to $H_{in} (T = R)$.  The system then thermalizes, and the probability of finding a state with localized energy 
$ E \ll R$ is $e^{ - E R}$. This state will, most probably, be a black hole of energy $E$, located at the origin of a static coordinate system.  The Schwarzschild dS black hole metric indeed tells us that a black hole of energy $E$ is a constrained state of the fundamental DOF, whose entropy is $R^2$.

\section{The Firewall Paradox}

As emphasized by Marolf\cite{marolf}, the essence of the firewall paradox is that QUEFT attributes only a single low energy state, the local Minkowski vacuum, to a black hole and explains the thermal nature of the outgoing radiation in terms of entanglement of the outgoing particles with particles near the horizon and inside it.  Retrieval of the black hole information by the external observer means that the outgoing particles are entangled with the distant detector, but monogamy of entanglement means that the short wavelength field theory DOF near the horizon are no longer in a state that resembles the local the Minkowski vacuum so that in-falling geodesic detectors will register high energy particles.   

In HST, there are independent Hamiltonians for different time-like trajectories, with consistency conditions connecting them.  For any geodesic, most of the states in the Hilbert space of a causal diamond of size $R_S$ contribute to terms in the Hamiltonian bounded by $1/R_S$.  Thus, any causal diamond, including that describing the region inside the stretched horizon\cite{membrane}\cite{susskinduglum} of a black hole, has a large number of low energy states, which are invisible to quantum field theory.  QUEFT describes only constrained states of the diamond, in which there are particles inside the black hole that can remain decoupled from the horizon DOF for times of order $R_S$.  

 An in-falling detector and particles that fall in with it, all localized objects, increase the size of the black hole Hilbert space, but as they fall through the horizon, the enlarged system begins in a constrained state in which the detector DOF are decoupled from those of the horizon.  The destruction of the detector at the singularity is simply the process of equilibration of detector and horizon DOF.  Hawking radiation is a Poincare recurrence of the thermalized black hole, in which a small subset of DOF, representing a Hawking particle, are decoupled from the rest by a random fluctuation.

The apparent singularity in the interior metric of the black hole is attributed to two features of the description in the previous paragraph\cite{gravres14}\cite{fw3}.  If I throw a second detector into the black hole after a scrambling time, then its experience is identical to that of the first, except for the increase in black hole mass due to the infall of the first detector.  That is, it must encounter a non-singular region of space-time for a period of order $R_S$ after falling through the horizon.  In a hydrodynamic space-time picture\cite{ted} of this phenomenon, the singularity encountered by the first detector must be causally separated from this region, which means that the interior space ``expands away from the horizon faster than the speed of light".  On the other hand, this detector too must eventually encounter a singularity indicating its equilibration with the horizon and the end of its career as a localized excitation.  Thus, the holoscreens of causal diamonds beginning at shorter and shorter time periods before it comes into equilibrium, must be shrinking.  Thus, the HST models can accommodate all of the qualitative properties of black holes, with no firewalls, in a framework that preserves locality, causality, unitarity, and much of the geometric structure of Minkowski space-time.  The consistency constraints for trajectories with relative velocity, which imply Lorentz invariance, are not generally satisfied, but existing models have enough parameters that one may hope that {\it some} of these models satisfy them.

\section{The Early Universe}

In cosmological models, we consider nested causal diamonds whose past tips all lie on a single space-like surface whose proper time is designated $t = 0$.  
$H_{in} (t) $ for very small $t$, acts on a Hilbert space of low dimension.  It is completely non-singular. At early times, we choose it, independently at each time, from a random distribution of Hamiltonians\footnote{We don't have space here to describe the details of this distribution.} in such a way that it scrambles the state of the system\cite{haydenstanford}.  We'll work strictly in $4$ space-time dimensions. As $t$ gets large, we insist that the Hamiltonian approach the $L_0$ generator of a $1 + 1$ dimensional chiral CFT, with central charge $\sim t^2$ .  The CFT has a UV cutoff $K$ and lives on an interval of length $D$, related by $KD \sim L$, where $L \gg 1$ is independent of $t$.  We'll argue that the space-time geometry corresponding to this model is a flat FRW metric, so that the volume of space on FRW slices scales like $t^3$.   Since we are in a generic state of the CFT the entropy scales like $t^2 KD$, while the energy is $ \sim t^2 K^2 D $
The entropy density on FRW slices is thus $\sim t^{-1} L $, while the energy density is $\sim\frac{K}{t} L$. We know from \cite{fsb} that an FRW cosmology that saturates the covariant entropy bound must have $p = \rho$, and this is achieved if we take $K \sim t^{-1}$.  Note that this is the only scaling compatible with the general FRW result that the energy density $\rho \sim t^{-2}$.  

The fact that our model actually represents an FRW space-time follows from the fact that we can make overlap rules for a collection of systems corresponding to other time-like trajectories.  Consider a sprinkling of time-like trajectories whose topology on the initial value surface is that of a regular 3 dimensional lattice.  Assign each of those trajectories the same sequence of time dependent Hamiltonians, and insist that they have the same initial state.  If two trajectories are separated by a minimum lattice walk of $S$ steps then we define the overlap Hilbert space at time $t$ to be the tensor factor of the Hilbert space on which $H_{in} (t - S)$ acts.  The consistency conditions are satisfied\footnote{We do not even have to make a unitary change of basis.  This is related to the fact that the current model maximizes the entropy at all times, and has no local excitations, which could be used to define reference states in Hilbert space.  A related fact is that nothing in the model depends on the choice of $H_{out} (t)$.}.  The overlap rules obey both homogeneity and isotropy.  The surface of points on the lattice which are a fixed number of steps away from a given point has the topology of a two sphere, and in the emergent metrical geometry defined by the causal relations, all of those points are the same space-like distance away.  In addition, the detailed models we have constructed have $SU(2)$ invariant Hamiltonians with the fast scrambling property\cite{sekinosusskind}, so that measurements made by a detector following any trajectory are rotationally invariant, at least when averaged over a few Planck times.  

The FRW metric expands forever, and cannot be positively curved.  Negatively curved FRW geometries do not saturate the covariant entropy bound even with equation of state $p = \rho$.  However, the cleanest argument that the spatial geometry is flat is that the model becomes scale invariant in the large $t$ limit where geometry/hydrodynamics is a good description.  Flat FRW models, with single component equations of state like $p = \rho$ have a conformal Killing symmetry, but curved ones do not.  There are a variety of other crude geometrical checks, which show that the model really does behave like the flat $p = \rho$ FRW geometry. Note that the singularity in the FRW geometry is completely spurious.  The geometric description is valid only for $t \gg 1$ in Planck units, the high entropy limit where hydrodynamic approximations become valid.   

If we insist that the Hilbert space reached at infinite proper time is finite dimensional, we get a model of a space-time with both $p = \rho$ and $p = - \rho$ contributions to its stress tensor\footnote{The overlap rules are also changed, to reflect the causal disconnection of trajectories that are outside each other's cosmological horizon.}.  The metric is flat FRW with
\begin{equation} a(t) = \sinh^{1/3} (3 t / R) , \end{equation} where $R$ is the ultimate dS radius.
There are two reasons why this model looks different at large $t$ from the model of eternal dS space discussed in the first section.  The first is purely kinematic.  We've described the Hamiltonian of our cosmology in terms of a time coordinate, which coincides with FRW time at the position of the time-like trajectory. The model of eternal dS space as a cutoff version of Minkowski space used the Hamiltonian for a time coordinate that coincides with static dS time.  The transformation between these coordinates, at large FRW time is a conformal transformation on the cutoff $1 + 1$ CFT, which rescales $K$ and $D$ so that $\langle H(R) \rangle \sim 1/R $.  

A much more important difference is that in the model of eternal dS space, we introduced, by hand, boundary conditions at $ t = - R$, corresponding to a finite set of jets.   If we start from any state of the model, {\it even a finite set of jets} at $t = - \infty$, then the probability of having the finite set of jets at $ t = - R$ is very small.  It's suppressed by a factor
\begin{equation} P = e^{ - 2 \pi E_{jets} R - \sum \pi R_i^2 M_P^2 }, \end{equation} where $E_{jets}$ is the total jet energy and $R_i$ is the Schwarzschild radius of a black hole whose energy is that of the $i$th jet.  In our cosmological model the initial state at the beginning of the dS era, is determined by the chaotic evolution during the $p = \rho$ era, and is quite generic.  It has only a tiny probability $P$ to be a jet state. Indeed, the probability to be any kind of localized excitation is
\begin{equation} P_{loc} = e^{ - 2\pi E R} . \end{equation}  Given a total energy $E$\footnote{The energy in these formulae is energy defined by the eternal dS space of radius $R$, as measured along a particular time-like geodesic.  } the most likely localized configuration is a single black hole.  Such an initial condition can never lead to a conventional radiation dominated universe, let alone one followed by a conventional matter dominated era, galaxies {\it etc.} .  Radiation, almost all of it massless particles, dribbles out and flows through the horizon, equilibrating with the DOF there.  The black hole ends its life in a high energy explosion, with a tiny fraction of the total energy of the original black hole.

Inhomogeneous configurations of black holes on FRW time slices, will suffer a similar fate.  They will collapse and coalesce into large black holes in a time scale much shorter than their evaporation time.  It is extremely plausible that the most likely way to get a universe with any sort of conventional homogeneous isotropic gas of particles, is to impose an initial condition that at trajectory times $t > t_I$ the initial state inside the horizon of a given trajectory is, for a period we will determine, given by an approximately homogeneous isotropic gas of black holes of horizon radius $t_I$ and density $< t_I^{-3} $. One can show \cite{holoinflation3} that such a gas will decay into radiation before the black holes coalesce.

Note that for $t < t_I$ this picture is consistent with our maximal entropy $p = \rho$ system inside the horizon as well as out. We can think of the full time dependent Hamiltonian as a sum of identical terms, each acting in a tensor factor Hilbert space of entropy $t_I^2$.  For $t > t_I$, the system inside the horizon couples together more DOF, but in a constrained state that describes the black hole gas.  As the horizon expands, in order to be consistent with the inside ``observer" who discovers more independent black holes, the Hamiltonian $H_{out} (t)$ must still treat individual systems of entropy $t_I^2$ as independent.  However, this is precisely the description of a collection of independent dS horizon volumes with dS radius $t_I$.  So our black hole gas inside the horizon implies an inflationary era outside the horizon.   In order for this to be consistent with the picture of the universe from other trajectories, it must be that {\it every} trajectory undergoes a period of inflation where $H_{in} (t)$ is a constant fast scrambling Hamiltonian on a Hilbert space of entropy $n^2$.  

In order to understand the picture outlined in the previous paragraph, we have to pay attention to the fact that the time slices of the proper time evolution for a given trajectory coincide with those of the FRW metric only at one space-point, the point where the trajectory penetrates the FRW time slice.  If we follow the convention that FRW time slices are horizontal, and draw the backward light cones from each point on the trajectory, then proper time slices are space-like hyperboloids lying between consecutive light cone boundaries. Looking at such a picture, we can see that, given the FRW metric which describes the hydrodynamics of the quantum state we've postulated, inflation must last for a conformal time $\eta_0 /2 $, where $\eta_0$ is the position of the pole in the scale factor $a( \eta )$, which signals the final dS state of the universe.

It is in fact impossible to have an exactly homogeneous and isotropic black hole gas, since black holes are finite chaotic quantum systems.  The mass and angular momentum of the black hole are thermodynamic averages and there are unavoidable fluctuations in both which are of order one over the square root of the entropy or $t_I^{-1}$ .  The precise ratio between the size of the mass and angular momentum fluctuations can in principle be calculated from a more detailed quantum model of the black hole, but this has not yet been done.  These fluctuations are fluctuations in the helicity zero and two components of the Weyl tensor and so are precisely the modes of metric fluctuation that are measured in the CMB (we work in comoving gauge where the stress tensor of the cosmic fluid remains homogeneous and isotropic).  The conventional measure $\zeta$ of scalar fluctuations is related to the scalar curvature fluctuation by a factor of $\epsilon^{-1}$ the slow roll parameter, so the ratio of scalar to tensor fluctuations is even larger than in field theory models of inflation. However, this can be masked by the fact that we can choose different slow roll metrics in the two models.  Thus, the most we can say is that the HST model can be made compatible with the data on two point functions, but suggests that tensor fluctuations will be rather small.

The spatial distribution of these fluctuations has an approximate $SO(1,4)$ invariance.  This comes from the fact that each inflationary horizon volume is described by a Hamiltonian that is approximately $SO(1,2)$ invariant, and the Hamiltonian outside the horizon is a sum of the independent $L_0$ generators.
When inflationary horizon volumes come into the horizon of the post-inflationary trajectory as localized black holes, they are also acted on by the $SO(3)$ rotation symmetry of $H_{in}$, so the sum is over $L_0 $ generators localized at angles
\begin{equation} H_{out} = \sum_i L_0 ({\bf Omega_i }), \end{equation} where ${\bf Omega_i}$ are unit $3$ vectors.  Defining
\begin{equation} J_{\pm a} = \sum_i \Omega_i^a L_{\pm} ({\bf \Omega_i}) , \end{equation} and identifying $H_{out}$ with $J_{04}$ we get an algebra that would converge to that of $SO(1,4)$ if the $\Omega_i$ were dense on the sphere.  Their density is exponential in the number of e-folds of inflation, so we closely approximate this algebra.  The spatial distribution of fluctuations therefore obeys the constraints of $SO(1,4)$ invariance, and this is enough to fit all extant data\cite{holoinflation2}.  

The HST model {\it does} make predictions that are different than field theory models for the tensor fluctuations, but we will have to wait till those are measured with some precision before we are able to differentiate between them.  Conceptually, the HST models are preferable because they are complete quantum mechanical models, with no singularities, and make only the minimal amount of fine tuning of initial conditions necessary to produce a universe with localized excitations more complicated than black holes.  Science however, is about data and experimental verification.  The arguments of \cite{holoinflation2} show that all extant cosmological data can be understood in a model with slow roll inflation, and scalar and tensor fluctuations approximately subject to the constraints of $SO(1,4)$ invariance.  Models as conceptually different as HST and quantum field theory can all satisfy these constraints, so it is clear that cosmological data do not afford us a sufficient handle on models of inflation.
The tensor fluctuations, particularly their three point function, could be a crucial discriminant between models, but in the HST model, they are predicted to be small and hard to measure.

\section{Conclusions}

HST realizes bulk localized excitations, both particles and black holes, as constrained states of DOF which live on the horizons of causal diamonds.  Horizon DOF account for gravitational entropy, and this resolves the firewall paradox, and gives a parametric explanation of the temperature of horizons, the form of gravitational scattering amplitudes at large impact parameter, as a function of energy and impact parameter, and the critical energy/impact parameter regime in which particle scattering leads to black hole formation.
HST also gives a novel explanation for the quasi-local nature of interactions (the time dependence of the HST Hamiltonian is crucial to this explanation). 
No extant models satisfy the consistency conditions of HST for trajectories with relative velocity, and this means that the models don't give Lorentz invariant S-matrices.  The existing models have enough parameters that one may hope to obtain Lorentz invariance by fine tuning.

HST also provides us with a completely finite model of early universe cosmology, from the ``initial singularity" through the reheating period after the end of inflation.  The model explains homogeneity, isotropy and flatness, and gives an account of CMB fluctuations more constrained than that of conventional inflation, and with less fine tuning.  Indeed, the fine tuning of initial conditions is required in order to explain why the universe ever contains anything more than a few large black holes in dS space.  The HST model computes (in principle) all properties of cosmology in terms of the slow roll metric and two other parameters, the c.c. and the post-inflationary density of an almost uniform gas of black holes, which decays into the Hot Big Bang. The predictions for 2 point scalar fluctuations are theoretically quite different from those of a field theory inflation model, but the difference can be masked by the freedom of choice of the slow roll metric. The predictions for tensor fluctuations have a (slightly) different power law, so if we can measure them over a large enough range of wavelengths we will be able to distinguish between the two theories.  The primordial tensor fluctuations in the HST model may be very small, depending on the details of the slow roll metric, but the gravitational waves from primordial black hole decay are suppressed only by a factor of one over the effective number of massless particles and have a tilt distinctly different from the standard slow roll inflation prediction. Tensor three point functions are VERY different in the two classes of model, but are probably too small to be measured in the forseeable future.

\vskip.3in
\begin{center}
{\bf Acknowledgments }
\end{center}
 The work of T.B. was supported in part by the Department of Energy.  I'd like to thank Willy Fischler for collaboration on the work described in this essay and for commenting on the manuscript. Sean Carroll and Michael Dine also read the manuscript and gave me important advice.  Finally, I'd like to thank the Physics Dept. at Georgia Tech, and particularly Predrag Cvitanovic and David Finkelstein, for hosting my sabbatical quarter during the period this essay was written.

\end{document}